%% file: manuscript.tex
\documentclass[journal=jctc, manuscript=article]{achemso}

\usepackage[utf8]{inputenc}
\usepackage{amsmath}
\usepackage{amsthm}
\usepackage{amssymb}
\usepackage{graphicx}
\usepackage{physics}
\graphicspath{{img/}}
 \DeclareGraphicsExtensions{.png,.svg}
\usepackage{braket}
\usepackage{todonotes}
\usepackage{booktabs}
\usepackage[nohyperlinks,nolist]{acronym}
\usepackage{siunitx}
\usepackage{chemformula}
\usepackage{acronym}
\usepackage{caption}
\usepackage{slashed}
\usepackage{subcaption}
\usepackage{here}
\usepackage[version=4]{mhchem}
\usepackage[capitalize,
            noabbrev]
            {cleveref}
            
\input{newcommands}

\title{The X2C of gold: X2C-inspired approximation of the Dirac coupling operator in a multiwavelet basis}

\newcommand{\uit}{UiT The Arctic University of Norway, 9037 Tromsø, Norway}
\newcommand{\hylleraas}{Hylleraas Centre for Quantum Molecular Sciences, \uit}
\newcommand{\chem}{Department of Chemistry, \uit}

\author{Quentin Pitteloud}
\affiliation{\hylleraas}
\alsoaffiliation{\chem}
\email{quentin.pitteloud@uit.no}

\author{Jacopo Masotti}
\affiliation{\hylleraas}
\alsoaffiliation{\chem}

\author{Luca Frediani}
\affiliation{\hylleraas}
\alsoaffiliation{\chem}
\email{luca.frediani@uit.no}

\input{acronyms.tex}

\begin{document}

\maketitle  
\begin{abstract}
We present a method to obtain the proven \textit{exact 2-component} (X2C) method, normally applied to Gaussian-type fixed Hilbert bases, in an adaptative non-uniform grid basis of the Hilbert space, namely Multiwavelets.
The X2C method manages to block-diagonalise the Dirac Hamiltonian restricted to the subspace of the basis, in an energy-independent manner.
The method takes inspiration from the success of the atomic mean-field X2C method on the Gaussian bases to represent the coupling operator as a projector onto the small components of the eigenstates of the constituent atoms of the system.

\end{abstract}


\newpage

\subsection{Introduction: Exact 2-component methods in finite bases} \label{sec:x2c_gaussian}

Compared to the Schrödinger equation solutions, Dirac spinors require 4 times as much information to be computed and represented.
The Hilbert space of Dirac spinors, $L^2(\mathbb{R}^3, \mathbb{C}^4)$, now encodes solutions of both positive and negative energy. 
Within bound state quantum mechanics, only the positive energy solutions are of interest, negative energy solutions corresponding to scattering states.
Fortunately, it is possible to block-diagonalise the Dirac Hamiltonian through an operator $\hat{X}$ and decouple positive (so-called \textit{large} components, denoted $\psi^L \in L^2(\mathbb{R}^3, \mathbb{C}^2)$) and negative energy solutions (\textit{small} components, $\psi^S\in L^2(\mathbb{R}^3, \mathbb{C}^2)$),
\begin{equation}\label{eq:exact_coupling_operator_relation}
    \hat{X}\psi^L = \psi^S, \ \Psi^{Dirac} = \begin{pmatrix}
        \psi^L \\ 
        \psi^S
    \end{pmatrix}.
\end{equation}
From the elimination of the small component approach, the coupling operator is expressed as
\begin{equation}\label{eq:exact_coupling_operator}
    \hat{X}(E) = \frac{c}{2mc^2+E-\hat{V}}(\boldsymbol{\sigma}\cdot\hat{\boldsymbol{p}}),
\end{equation}
where $E$ is the energy eigenvalue of the considered solution, $\boldsymbol{\sigma} = (\sigma_1,\sigma_2,\sigma_3)$ the vector of Pauli matrices.
Alternatively, we can recover the coupling operator from solutions of the Dirac equation.
For simplicity, let us restrict our considerations to the subspace of bound solutions of the Dirac Hamiltonian within an external potential.
Assuming we have the complete set of bound solutions $\{\ket{\Psi_i}\}_{i\in \mathbb{N}}$,  the exact coupling operator can be written as 
\begin{equation}
    \hat{X} = \sum_{i,j\in \mathbb{N}} \ket{\psi^S_i}\big(S^L\big)_{ij}^{-1} \bra{\psi^L_j}, \ \ket{\Psi_i} = \left|\begin{pmatrix}
        \psi^L_i \\
        \psi^S_i
    \end{pmatrix}\right\rangle,
\end{equation}
where $S^L_{ij} := \braket{\psi^L_i}{\psi^L_j}$ is the overlap matrix of the large components, and so 
\begin{equation}
    \big(S^L\big)_{ij}^{-1} \bra{\psi^L_j} = \big(S^L\big)_{ij}^{-1}\big(\ket{\psi^L_j}\big)^\dagger =: \ket{\psi^L_i}^+, 
\end{equation}
is the Moore-Penrose pseudoinverse.


Discretising the Hilbert space by choosing a (finite) basis $\{\phi_i\}_{i\in [\![N]\!]}$, it is possible to write the restriction to the basis' span of this operator as a matrix with no explicit dependence on the energy.
Indeed, denoting the coefficients vectors in the chosen basis of the large and small components as $\psi^{L,S} = C^{L,S}_{i}\phi_i, \  C^{L,S}\in \mathbb{C}^{2N}$, there exists a matrix $R$ such that 
\begin{equation} \label{eq:coupling_matrix_representation}
    RC^L = C^S, \ R \in \mathbb{M}_{2N\times 2N} (\mathbb{C}).
\end{equation}
Comparing the matrix equation \ref{eq:coupling_matrix_representation} to the coupling operator action (equation \ref{eq:exact_coupling_operator_relation}), we see that the matrix $R$ represents the action of the coupling operator within the subspace spanned by the Hilbert space basis.
Furthermore, we can compute the matrix $R$ \textit{a priori}, by considering the Dirac equation restricted to the basis' span so that we can exploit the coupling matrix equation \ref{eq:coupling_matrix_representation},
\begin{align}
    &\begin{pmatrix}
        \hat{V}\psi^L & c(\boldsymbol{\sigma} \cdot \boldsymbol{p} )\psi^S   \\
    c(\boldsymbol{\sigma} \cdot \boldsymbol{p} )\psi^L  & \hat{V}\psi^S  
    \end{pmatrix} = \begin{pmatrix}
         E\psi^L \\
         E\psi^S 
    \end{pmatrix}\\ 
    \Leftrightarrow &\begin{cases}
        \hat{V}C^L\phi + c(\boldsymbol{\sigma} \cdot \boldsymbol{p} )C^S\phi =   EC^L\phi \\
    c(\boldsymbol{\sigma} \cdot \boldsymbol{p} )C^L\phi  + \hat{V}C^S \phi = EC^S\phi
    \end{cases} \\
    \Leftrightarrow &\begin{cases}
        VC^L + \Pi RC^L =   EC^L  \\
    \Pi C^L  + VRC^L  = ERC^L
    \end{cases} \\
    \Leftrightarrow &\begin{cases}
        RVC^L + R\Pi RC^L =   ERC^L  \\
    \Pi C^L  + VRC^L  = ERC^L
    \end{cases}\\
    \Leftrightarrow & RVC^L + R\Pi RC^L = \Pi C^L  + VRC^L, \label{eq:Ricatti_x2c}
\end{align}
we arrive at a quadratic matrix equation, the algebraic Riccati equation, where we defined the momentum matrix representation as $\Pi_{ij} = \bra{\phi_i}c(\boldsymbol{\sigma} \cdot \boldsymbol{p} )\ket{\phi_j}$ and similarly the interaction potential matrix $V_{ij} = \bra{\phi_i}\hat{V}\ket{\phi_j}$.
Solving this equation in practice is possible but may be prohibitively expensive for large systems due to the correspondingly large bases required \cite{laub_schur_1979, bunse-gerstner_computational_nodate}. 

By construction, the solution to the Riccati equation verifies the relation of equation \ref{eq:coupling_matrix_representation}, and given $2N$ set of Dirac solution we could recover $R$. 
While obviously useless as is, for complex systems that can be easily decomposed into parts (such as molcules into atoms), it serves as the basis for a very useful approximation.
The exact coupling operator (equation \ref{eq:exact_coupling_operator}) depends explicitly on the momentum of the particles, which by energy conservation will be strongest close to the source of the potential energy, such a nuclei in atoms and molecules. 
We infer that the matrix $R$ of \textit{e.g.} a molecule will be approximate to a very good extent by the direct sum of the coupling matrices $R^i$ of its constituent atoms, assuming that the discretisation of the molecule's Hilbert space is itself the direct sum of its atoms'.
To recover exactly the matrix $R_{\sigma}$ for each subsystem, we need to have a complete set of $2N$ linearly independent vectors to solve the matrix equation
\begin{equation} \label{eq:amfX2C_coupling_matrix}
    R_\sigma= \bar{C}_\sigma^S\big(\bar{C}_\sigma^L\big)^{-1}, 
\end{equation}
where $\bar{C}_\sigma^{S,L} \in \mathbb{M}_{2N\times 2N} (\mathbb{C})$ represent the matrices composed of the coefficient vectors $C^{S,L}\in \mathbb{C}^{2N}$ of the subsystem $\sigma$.
The set of vectors can naturally be selected as the physically relevant eigenvectors of the Hamiltonian matrix expressed in the basis.
In the common choice of Gaussian bases in quantum chemistry, diagonalising the Hamiltonian matrix produces approximations of the bound eigenstates of the Hamiltonian along with physically irrelevant states, so-called \textit{virtual} states.
Both physical and virtual states are identified by their corresponding energy eigenvalues, bound states having negative energy and virtual states positive. 
An important side note is that the virtual states do not sample or discretise the continuum in any way but in their energy eigenvalue being contained in the positive continuum spectrum.
Indeed, Gaussian functions, through their nature as integrable function, cannot represent scattering states (propagating waves).
However the virtual states are required for the exact computation of the matrix $R_s$, at first to provide full-rank coefficient matrices $\bar{C}_\sigma^{L,S}$.
Once the coupling matrices $R_s$ of each subsystem $\sigma$ is known, we can approximate the matrix $R$ of the full system as
\begin{equation}
    R = \bigoplus_{\sigma=1}^\Sigma R_\sigma,
\end{equation}
assuming that we express the basis of the full system as the union of the subsystems' bases. 
This assumption is quite reasonable for fixed basis computations, but such assumption can be relaxed through linear algebra, using change of basis matrices for example.
The construction of the $R$ matrix (equation \ref{eq:amfX2C_coupling_matrix}) forms the core idea of the \textit{amfX2C} method, pioneered by Knecht \textit{et al.}, though the authors refine the idea further within the context of Gaussian basis formalism \cite{knecht_mossbauer_2011}.
As a first step towards obtaining X2C methods within a multiwavelet framework, it is this core idea that we shall use a basis.

\section{Adapting amfX2C to multiwavelets}

Perhaps multiwavelets bases' greatest advantage is their adaptability, allowing for a highly refined grid where necessary without wasting resources describing empty or irrelevant space.
However, as a non-fixed-grid-based method, this feature also creates a challenge in the computation of the coupling matrix of the full system $R$.
The full system's basis cannot be assumed to be the direct sum of the subsystem's bases, and care will be needed to construct the representation of $\hat{X}$ within that basis.
This challenge would be surmountable, yet the relatively large amount of basis function typically involved in a multiwavelet basis render a direct expression of $R_\sigma$ into the basis directly impractical.
Instead, we shall piggyback on the success found by this method in Gaussian bases to represent the operator within a Gaussian basis.
Indeed, as we shall discuss later in section \ref{sec:Gaussian_success_X2C}, Gaussian functions strike a good balance between accuracy of the representation and reduced basis size.

On the other hand, the formalism we propose does not necessary require a Gaussian representation.
The cornerstone of the idea is the transition matrix expression of the coupling operator, that we recall here, 
\begin{equation}
    \hat{X} = \sum_{i,j\in \mathbb{N}} \ket{\psi^S_i}\big(S^L\big)_{ij}^{-1} \bra{\psi^L_j}, \ \ket{\Psi_i} = \left|\begin{pmatrix}
        \psi^L_i \\
        \psi^S_i
    \end{pmatrix}\right\rangle,
\end{equation}
where $S^L_{ij} := \braket{\psi^L_i}{\psi^L_j}$ is the overlap matrix of the large components.
We propose an approximation of the coupling operator constructed from a finite set of physically relevant vectors of the Hilbert space, such as the eigenstates of the isolated subsystems (atoms) of the total system (molecule or unit cell). 
Assuming that for each isolated atom we have obtained a set of physically relevant states 
\begin{equation}
    \{ \ket{\Phi_{\sigma,i}}\}_{\sigma\in[\![\Sigma]\!], i\in [\![N_\sigma]\!]}, \text{ s.t. } \hat{H}_\sigma \ket{\Phi_{\sigma,i}} = E_{\sigma,i}\ket{\Phi_{\sigma,i}} \text{ with } \ket{\Phi_{\sigma,i}} = \left|\begin{pmatrix}
        \phi^L_{\sigma,i} \\
        \phi^S_{\sigma,i}
    \end{pmatrix}\right\rangle,
\end{equation}
where $E_{\sigma,i}$ is an eigenvalue of the Hamiltonian (but does not have to hold a physical meaning), we can construct a projector onto $\text{Span}\{ \ket{\Phi_{\sigma,i}}\}$ as 
\begin{equation} \label{eq:approx_coupling_operator}
    \hat{Y} = \sum_{i,j = 1 }^{N_{tot}} \ket{\phi^S_{i}}\big(S^L\big)_{ij}^{-1} \bra{\phi^L_{j}}, 
\end{equation}
where we redefined the indices $i,j$ to run over all $N_{tot}$ elements of the union of the atomic bases, with $N_{tot}:=\sum_\sigma N_\sigma$.
Unfortunately, the eigenstates of the full (molecular) Hamiltonian cannot be assumed to be held within that subspace, nevertheless it can be assumed that the true eigenstates can be well approximated by linear combinations of the atomic eigenstates.
As we discussed in section \ref{sec:x2c_gaussian}, this claim is motivated by the momentum dependence of the coupling operator which focuses its effect around the nuclei, where the momentum of the orbiting electrons is highest, whereas the atomic eigenstates span a relatively large subspace of functions located close to atoms.
Of course, further motivation comes from the empirical evidence of the success of \textit{amfX2C} in Gaussian-type orbitals.

The accuracy of the projector $\hat{Y}$ depends on the amount of states are used to represent the small components. 
Within a multiwavelet basis, only negative energy (bound) eigenstates are easily computable, as this basis does not provide any description of virtual states. 
Within this basis, positive energy states are necessarily scattering states, posing their own challenge for representation. 
Instead as we stated earlier, we shall accept to lose some accuracy of the description of the small component by choosing a Gaussian-type representation of the atomic orbitals, as in equation \ref{eq:approx_coupling_operator}.
Expressed formally, all components $\alpha$ and $\beta$ of Large and Small spinors are expressed as polynomial-dressed Gaussian functions, that is for any $T = L,S$, $i\in[\![N_{tot}]\!]$, $\gamma = \alpha, \beta$
\begin{equation}
    \phi^{T, \gamma}_{i}(\boldsymbol{x}) = \sum_{j} c_{i, j}^{T, \gamma} \sum_p a_p P_p(x) e^{-b_p (\boldsymbol{x}-\boldsymbol{x}_p)^2},
\end{equation}
with $P_p$ a polynomial and $a_p\in \mathbb{C},\ b_p \in \mathbb{R}$ set by the choice of atomic basis, and $c_{i, j}^{T, \gamma} \in \mathbb{C}$ are expansion coefficients.
We then aim to represent this function in a multiwavelet basis, for which we first need to project the Gaussian basis functions.
As any other $L^2$ function, we can express the Gaussian basis functions in the multiwavelet basis, by computing the overlap between the bases.
Simplifying the notation to index the 4 components $\gamma =1,\dots,4$ of a Dirac spinors instead of $L,S$ and $\alpha,\beta$, the expansion in the multiscaling functions $\mathfrak{s}_{k,j}^n$ is 
\begin{equation}
    \langle \phi^{\gamma}_{i}, \mathfrak{s}_{k,\boldsymbol{y}}^n\rangle = \int_{\mathbb{R}^3} \dd^3 \boldsymbol{x} \ \phi^{\gamma}_{i}(\boldsymbol{x}) \mathfrak{s}_{k,\boldsymbol{y}}^n(\boldsymbol{x}) =: s_{k, i\boldsymbol{y}}^{n, \gamma},
\end{equation}
and similarly on the multiwavelets functions,
\begin{equation}
    \langle \phi^{\gamma}_{i}, \mathfrak{w}_{k,\boldsymbol{y}}^n\rangle = \int_{\mathbb{R}^3} \dd^3 \boldsymbol{x} \ \phi^{\gamma}_{i}(\boldsymbol{x}) \mathfrak{w}_{k,\boldsymbol{y}}^n(\boldsymbol{x}) =: w_{k, i\boldsymbol{y}}^{n, \gamma}.
\end{equation}
Computing these overlaps allow us to represent the Gaussian functions as mutliwavelets trees \cite{alpert_wavelet-like_1993, mallat_theory_nodate}. 
Writing $V_n$, $W_n$ the span of the scaling, respectively wavelet functions at scale $n$, the $L^2$ space can be written as the sum 
\begin{equation}
    L^2 = V_0 + \bigoplus^\infty_{n=0}W_n,
\end{equation}
and we can thus write, for $\epsilon >0$,
\begin{equation}
    \phi^{\gamma}_i(\boldsymbol{x}) =\sum_{k=0}^{K}\sum_{y_1, y_2,y_3=-\infty}^{\infty}\left(  s_{k, i\boldsymbol{y}}^{0, \gamma}  \mathfrak{s}_{k,\boldsymbol{y}}^0+ \sum_{j=0}^{\infty} w_{k, ij}^{n, \gamma} \mathfrak{w}_{k,\boldsymbol{y}}^n\right).
\end{equation}

In practice, the projector $\hat{Y}$ can simply be stored as the Gaussian-represented Dirac spinors from the atomic calculations, as its action on an arbitrary state simply gives a linear combination of the Gaussian approximated small components,
\begin{equation}
    \hat{Y}\ket{\xi} =  \sum_{i,j=1}^{N_{tot}} \ket{\phi^S_{i}} \big(S^L\big)_{ij}^{-1} \bra{\phi^L_{j}} \xi \rangle = \sum_{i=1}^{N_{tot}}A_{ i} \ket{\phi^S_{i}},
\end{equation}
where $A:= \big(S^L\big)^{-1} S^{L\xi}$, and $\big( S^{L\xi}\big)_{i} = \bra{\phi^L_{i}} \xi \rangle$.
In particular, given the fixed nature of the coupling operator approximation, we shall decide to represent the orthogonalised large components spinors to minimise redundant computations,
\begin{equation}
    \ket{\tilde{\phi}^L_i} := \sum_{j}^{N_{tot}}  \big(S^L\big)_{ij}^{-1} \ket{\phi^L_j},
\end{equation}
which simplifies the approximated coupling operator to 
\begin{equation}
    \hat{Y}=  \sum_{i=1}^{N_{tot}} \ket{\phi^S_{i}} \bra{\tilde{\phi}^L_{i}}.
\end{equation}

\section{Atomic Small Component Hamiltonian and propagator}

With the coupling operator now set, we shall derive the Hamiltonian and propagator required to solve its eigenvalue problem.

The Dirac Hamiltonian is 
\begin{equation}
    \hat{H}^D = c\alpha \cdot \hat{\boldsymbol{p}} + \beta mc^2 + \hat{V}.
\end{equation}
In matrix form, in the Dirac basis, we have
\begin{equation}
    \hat{H}^D = \begin{pmatrix}
        \hat{V}+mc^2 & c(\boldsymbol{\sigma} \cdot \boldsymbol{{\hat{p}}} )  \\
    c(\boldsymbol{\sigma} \cdot \boldsymbol{\hat{p}} )  &  \hat{V}-mc^2 
    \end{pmatrix},
\end{equation}
giving the eigenvalue equation 
\begin{equation} \label{eq:Dirac_eigval_matrix}
    \begin{pmatrix}
        \hat{V}+mc^2-E & c(\boldsymbol{\sigma} \cdot \boldsymbol{{\hat{p}}} )  \\
        c(\boldsymbol{\sigma} \cdot \boldsymbol{\hat{p}} )  &  \hat{V}-mc^2-E 
    \end{pmatrix}\ket{\Psi}=0.
\end{equation}
Given a coupling operator $\hat{X}$, not necessarily the exact operator, we can enforce the small component to verify $\ket{\psi^S} = \hat{X}\ket{\psi^L}$, giving us an expectation value of the Hamiltonian for an arbitrary state $\ket{\Psi} = (\ket{\psi^L}, \ket{\psi^S})^T$ as
\begin{align}
    \varepsilon =& \bra{\Psi}\hat{H}^D \ket{\Psi} \\
    =& \bra{\psi^L}(\hat{V}+mc^2)\ket{\psi^L} + \bra{\psi^S}(\hat{V}-mc^2)\ket{\psi^S} + \bra{\psi^L}c\sigma\cdot \hat{\boldsymbol{p}}\ket{\psi^S}+\bra{\psi^S}c\sigma\cdot \hat{\boldsymbol{p}}\ket{\psi^L} \\
    =& \bra{\psi^L}(\hat{V}+mc^2)\ket{\psi^L} + \bra{\psi^L}\hat{X}^\dagger(\hat{V}-mc^2)\hat{X}\ket{\psi^L} + \bra{\psi^L}c\sigma\cdot \hat{\boldsymbol{p}}\hat{X}\ket{\psi^L}+\bra{\psi^L}\hat{X}^\dagger c\sigma\cdot \hat{\boldsymbol{p}}\ket{\psi^L}\\
    =&\bra{\psi^L}\left(\hat{V}+mc^2 +\hat{X}^\dagger(\hat{V}-mc^2)\hat{X}+c\sigma\cdot \hat{\boldsymbol{p}}\hat{X}+\hat{X}^\dagger c\sigma\cdot \hat{\boldsymbol{p}}\right)\ket{\psi^L},
\end{align}
from which we extract the effective Hamiltonian 
\begin{equation}\label{eq:Hamiltonian_ASC}
    \hat{H}^{ASC} = \hat{V}+mc^2 +\hat{X}^\dagger(\hat{V}-mc^2)\hat{X}+c\sigma\cdot \hat{\boldsymbol{p}}\hat{X}+\hat{X}^\dagger c\sigma\cdot \hat{\boldsymbol{p}}.
\end{equation}
This formulation allows us to avoid the computation of the normalisation operator $N:= \sqrt{1+ \hat{X}^\dagger\hat{X}}$ and its inverse, as long as the set of large component spinors $\{\ket{\psi^L_i}\}_{i\in [\![N]]\!}$ are orthonormalised using the overlap of the full Dirac spinors.
Using the Löwdin orthonormalisation procedure, the overlap matrix to be considered is
\begin{equation} \label{eq:overlap_matrix_ASC}
    S = \bra{\Psi}{\Psi}\rangle = \bra{\psi^L}{\psi^L}\rangle + \bra{\psi^S}{\psi^S}\rangle = \bra{\psi^L}{\psi^L}\rangle+ \bra{\psi^L}\hat{X}^\dagger \hat{X}\ket{\psi^L}.
\end{equation}
Expressing the normalisation operator $\hat{N}$ within the projector approximation would in practice be expensive, at least with a naive Taylor expansion, as opposed to the comparatively cheap computation of the overlap matrix in equation \ref{eq:overlap_matrix_ASC}.

To solve the eigenvalue equation \ref{eq:Dirac_eigval_matrix}, we shall proceed to do it iteratively.
We shall have to express the eigenvalue problem in terms of the bound-state Helmholtz/Yukawa kernel. 
Starting from
\begin{equation}
    \hat{H}^D\ket{\Psi} = \left(c\alpha \cdot \hat{\boldsymbol{p}} + \beta mc^2 + \hat{V}\right)\ket{\Psi} = \varepsilon\ket{\Psi},
\end{equation}
we can reorder the problem as
\begin{equation}\label{eq:Dirac_eq_reordered}
    \left(c\alpha \cdot \hat{\boldsymbol{p}} + \beta mc^2 -\varepsilon\right)\ket{\Psi} = -\hat{V}\ket{\Psi}.
\end{equation}
Multiplying the left hand side by
\begin{equation}
    1 = \left(\hat{h}^D + \varepsilon\right)^{-1}\left(\hat{h}^D + \varepsilon\right), \text{ with } \hat{h}^D:=\left(c\alpha \cdot \hat{\boldsymbol{p}} + \beta mc^2\right), 
\end{equation}
in equation \ref{eq:Dirac_eq_reordered}, we find 
\begin{equation}\label{eq:Dirac_eq_reordered_square_trick}
    \left(\hat{h}^D + \varepsilon\right)^{-1} \left((\hat{h}^D)^2 - \varepsilon^2\right)\ket{\Psi} = -\hat{V}\ket{\Psi}.
\end{equation}
Using the anticommutation relations of the $\alpha_i, \beta$ variables, we know that 
\begin{equation}
    (\hat{h}^D)^2 = c^2\hat{\boldsymbol{p}}^2 +m^2c^4 = - c^2 \hbar^2\nabla^2  + m^2 c^4.
\end{equation}
We can rewrite equation \ref{eq:Dirac_eq_reordered_square_trick} as an integral equation using the Green's function associated with the operator $\left(\hat{h}^D + \varepsilon\right) \left(- c^2 \hbar^2\nabla^2  + m^2 c^4 - \varepsilon^2\right)^{-1}$, and writing it as $\left[\left(\hat{h}^D + \varepsilon\right)G_0^\mu\right]$, we find
\begin{equation}\label{eq:Dirac_integral}
     \ket{\Psi} = \left[\left(\hat{h}^D + \varepsilon\right)G_0^\mu\right] * \left( -\frac{1}{\hbar^2 c^2}\hat{V}\ket{\Psi} \right),
\end{equation}
where $G_0^\mu \sim (-\nabla^2 +\mu^2)^{-1}$ is the Yukawa kernel, and $\mu := \frac{1}{\hbar}\sqrt{m^2c^2 -\frac{\varepsilon^2}{c^2}}$.
Furthermore, following the works of Blackledge and Babajanov \cite{blackledge_dirac_2013}, the convolution of the Green's kernel allow us to permute the terms, and we can reorder equation \ref{eq:Dirac_integral} as
\begin{equation}
    \ket{\Psi} = G_0^\mu * \left( -\left[\hat{h}^D + \varepsilon\right]\hat{V}\ket{\Psi}\right).
\end{equation}
Expanding it again in its matrix form, we find
\begin{align}
    \ket{\Psi} &= G_0^\mu * \left[ -\frac{1}{\hbar^2 c^2}\begin{pmatrix}
        \varepsilon+mc^2 & c\sigma\cdot \hat{\boldsymbol{p}} \\
        c\sigma\cdot \hat{\boldsymbol{p}} & \varepsilon-mc^2 
    \end{pmatrix} \begin{pmatrix}
        \hat{V}\ket{\psi^L }\\
        \hat{V}\hat{X}\ket{\psi^L }
    \end{pmatrix}\right]\\
    &= G_0^\mu * \left[ -\frac{1}{\hbar^2 c^2}\begin{pmatrix}
        (\varepsilon+mc^2)\hat{V}\ket{\psi^L }\  +\quad  c\sigma\cdot \hat{\boldsymbol{p}}\hat{V}\hat{X}\ket{\psi^L } \\
        c\sigma\cdot \hat{\boldsymbol{p}}\hat{V}\ket{\psi^L }\quad +\  (\varepsilon-mc^2 )\hat{V}\hat{X}\ket{\psi^L }
    \end{pmatrix}\right].
\end{align}
The Green's function convolution acts diagonally in the spin space, and we are only interested in the evolution of the large component $\ket{\psi^L}$, we can then write 
\begin{equation}
    \ket{\psi^L}= -G_0^\mu * \left[\frac{1}{\hbar^2 c^2}\left(
        (\varepsilon+mc^2)\hat{V}\  +\  c\sigma\cdot \hat{\boldsymbol{p}}\hat{V}\hat{X}\right)\ket{\psi^L }
    \right].
\end{equation}
\newpage

\section{Advantages and limitations of the method}
Approximating the coupling operator as a projector manner makes it evident that the small components will be forced to keep their Gaussian representation, no matter the basis used for the large components.
However, as the latter are represented in a larger, more complete basis, we will be able to sample some of the information lost by the incompleteness of the Gaussian basis.
Writing Large/Small components spinors $\ket{\psi^{L,S}}$ of an arbitrary solution the eigenvalue equation of Hamiltonian \ref{eq:Hamiltonian_ASC} as the Gaussian span projections $\ket{\phi^{L,S}}$ and their respective orthogonal complement $\ket{\zeta^{L,S}}$, 
\begin{equation}
    \ket{\psi^{L,S}} = \ket{\phi^{L,S}} + \ket{\zeta^{L,S}}, \ \bra{\phi^{L,S}}{\zeta^{L,S}}\rangle = 0,
\end{equation}
we can write the Hamiltonian expectation values as
\begin{align}
    \bra{\psi^{L}}\hat{H}^{ASC}\ket{\psi^{L}} =& \left(\bra{\phi^{L}}+\bra{\zeta^{L}}\right)\hat{H}^{ASC}\left(\ket{\phi^{L}}+\ket{\zeta^{L}}\right) \\
    =& \bra{\psi^L}\left(\hat{V}+mc^2\right)\ket{\psi^L} +\bra{\phi^S}\left(\hat{V}+mc^2\right)\ket{\phi^S} + \bra{\psi^L}c\boldsymbol{\sigma}\cdot \hat{\boldsymbol{p}}\ket{\phi^S} + \bra{\phi^S}c\boldsymbol{\sigma}\cdot \hat{\boldsymbol{p}}\ket{\psi^L} \nonumber \\
    =& \bra{\phi^L}\left(\hat{V}+mc^2\right)\ket{\phi^L} +\bra{\phi^S}\left(\hat{V}+mc^2\right)\ket{\phi^S} + \bra{\phi^L}c\boldsymbol{\sigma}\cdot \hat{\boldsymbol{p}}\ket{\phi^S} + \bra{\phi^S}c\boldsymbol{\sigma}\cdot \hat{\boldsymbol{p}}\ket{\phi^L} \nonumber \\
    & \  + \bra{\zeta^L}\left(\hat{V}+mc^2\right)\ket{\zeta^L} +\bra{\zeta^L}c\boldsymbol{\sigma}\cdot \hat{\boldsymbol{p}}\ket{\phi^S} + \bra{\phi^S}c\boldsymbol{\sigma}\cdot \hat{\boldsymbol{p}}\ket{\zeta^L} \nonumber\\
    =&  \varepsilon_{Gaussian}  + \bra{\zeta^L}\left(\hat{V}+mc^2\right)\ket{\zeta^L} +\bra{\zeta^L}c\boldsymbol{\sigma}\cdot \hat{\boldsymbol{p}}\ket{\phi^S} + \bra{\phi^S}c\boldsymbol{\sigma}\cdot \hat{\boldsymbol{p}}\ket{\zeta^L}, \nonumber
\end{align}
with $\varepsilon_{Gaussian}$ the expectation value obtainable from Gaussian functions.
Of course, in practice, the orthogonal complement basis, $\ket{\zeta^{L}}$, will be incomplete.
However, multiwavelets' adaptativity property ensures that the basis is complete within a controllable numerical threshold, and with a tight enough threshold parameter, we can sample the orthogonal space.

%

On the other hand, increasing the basis size of the large component while keeping the small component at a low(er) precision can cause a violation of the kinetic balance (KB) \cite{stanton_kinetic_1984}.
In moderate cases, it translates into inaccurate results and in more serious cases complete variational breakdown.
Schematically, this is caused by the high level of description of the large component allowing the wavefunction to contract down to the nuclear core as the small components' limited description and contribution cannot counteract.
The kinetic imbalance can originate from the Gaussian small component undersampling the small component Hilbert space, preventing the kinetic balance condition,
\begin{equation}
    \left(\hat{V}-E \right)\ket{\Psi_i^L} = c\sigma\cdot \hat{\boldsymbol{p}}\ket{\Psi_i^S},
\end{equation}
from being verified.

Preliminary tests show that the naive approach of a static Gaussian small component creates a small error on the energy. 
For a hydrogenic Gold cation Au$^{78+}$, a relative error of $4.95\cdot10^{-4}$ has been observed when using the Dyall.v2z basis for the small component.

\begin{acknowledgement}
This work was supported by the Norwegian Research Council through a Centre of Excellence grant (Hylleraas Centre 262695), a FRIPRO grant (ReMRChem 324590), and by UNINETT Sigma2 through grants of computer time (nn9330k and nn14654k). 
\end{acknowledgement}

We use the CRediT taxonomy of contributor roles.\cite{Allen2014-rd,Brand2015-qc}
The ``Investigation'' role also includes the ``Methodology'', ``Software'', and ``Validation'' roles.
The ``Analysis'' role also includes the ``Formal analysis'' and ``Visualization''
roles. The ``Funding acquisition'' role also includes the ``Resources'' role.
We visualize contributor roles in the following authorship attribution matrix,

\begin{table}[ht]
\caption{Levels of contribution: \textcolor{blue}{major}, \textcolor{cyan}{minor}.}
\label{tbl:contribs}
\begin{tabular}{lcccc}
\hline\hline
                              &  QP    & JM     & LF     \\ \hline
    Conceptualization         & \major & \minor   &     \minor            \\ 
Investigation             & \major &   &  & \\ 
    Data curation             & \major &    &  \\ 
    Supervision               &  \major      &    & \major \\ 
    Writing -- original draft & \major & \minor  & \minor \\
    Writing -- revisions      & \major &  \minor  & \major \\
    Funding acquisition       &        &         & \major \\ 
    Project administration    &        &   & \major \\ \hline\hline
\end{tabular}
\end{table}

\bibliography{X2C}

\end{document}

%% file: newcommands.tex
\makeatletter
\renewcommand{\@cite}[1]{
{$\!\! ^{(#1)}$}}
\makeatother

\newcommand{\major}{\colorbox{blue}{\qquad}}
\newcommand{\minor}{\colorbox{cyan}{\qquad}}



%% file: acronyms.tex
\begin{acronym}
\acro{AO}{atomic orbital}
\acro{API}{Application Programmer Interface}
\acro{AUS}{Advanced User Support}
\acro{BO}{Born-Oppenheimer}  
\acro{CBS}{complete basis set}
\acro{CC}{Coupled Cluster}
\acro{CTCC}{Centre for Theoretical and Computational Chemistry}
\acro{CoE}{Centre of Excellence}
\acro{DC}{dielectric continuum}  
\acro{DFT}{density functional theory}  
\acro{DKH}{Douglas-Kroll-Hess}
\acro{ESC}{Elimination of Small Components}
\acro{EFP}{effective fragment potential}
\acro{ECP}{effective core potential}
\acro{EU}{European Union}
\acro{FW}{Foldy–Wouthuysen}
\acro{GGA}{generalized gradient approximation}
\acro{GPE}{Generalized Poisson Equation}
\acro{GTO}{Gaussian type orbital}
\acro{HF}{Hartree-Fock}  
\acro{HPC}{high-performance computing}
\acro{Hylleraas}[HC]{Hylleraas Centre for Quantum Molecular Sciences}
\acro{IEF}{Integral Equation Formalism}
\acro{IGLO}{individual gauge for localized orbitals}
\acro{KB}{kinetic balance}
\acro{KS}{Kohn-Sham}
\acro{LAO}{London atomic orbital}
\acro{LAPW}{linearized augmented plane wave}
\acro{LDA}{local density approximation}
\acro{MAD}{mean absolute deviation}
\acro{maxAD}{maximum absolute deviation}
\acro{MM}{molecular mechanics}  
\acro{MCSCF}{multiconfiguration self consistent field}
\acro{MPA}{multiphoton absorption}
\acro{MRA}{multiresolution analysis}
\acro{MW}{multiwavelet}
\acro{NAO}{numerical atomic orbital}
\acro{NeIC}{nordic e-infrastructure collaboration}
\acro{NMR}{nuclear magnetic resonance}
\acro{NP}{nanoparticle}  
\acro{OLED}{organic light emitting diode}
\acro{PAW}{projector augmented wave}
\acro{PBC}{Periodic Boundary Condition}
\acro{PCM}{polarizable continuum model}
\acro{PW}{plane wave}
\acro{QC}{quantum chemistry}  
\acro{QM/MM}{quantum mechanics/molecular mechanics}  
\acro{QM}{quantum mechanics}
\acro{RA}{Regular Approximation}
\acro{RCN}{Research Council of Norway}
\acro{RMSD}{root mean square deviation}
\acro{RKB}{restricted kinetic balance}
\acro{SC}{semiconductor}
\acro{SCF}{Self Consistent Field}
\acro{STSM}{short-term scientific mission}
\acro{SAPT}{symmetry-adapted perturbation theory}
\acro{SERS}{surface-enhanced raman scattering}
\acro{WPREL}[WP1]{Work Package 1}
\acro{WPROP}[WP2]{Work Package 2}
\acro{WPAPP}[WP3]{Work Package 3}
\acro{WP}{Work Package}  
\acro{X2C}{exact two-component}
\acro{ZORA}{zero-order regular approximation}
\acro{ae}{almost everywhere}
\acro{BVP}{boundary value problem}
\acro{PDE}{partial differential equation}
\acro{RDM}{1-body reduced density matrix}
\end{acronym}